\documentstyle[epsfig,twocolumn]{article} 
\title{ \vspace{-2.4cm} 
\hspace*{11cm} {\large KEK-Preprint 98-163} \\
\hspace*{12.5cm} {\large October 1998} \\
  \vspace{0.3cm} {\Large\bf
GAMMA-GAMMA, GAMMA-ELECTRON COLLIDERS} \thanks{Invited talk at XII
Intern. conf. on High Energy Accelerators (HEACC'98), Sept.7-12, 1998,
Dubna, Russia}} \author{Valery Telnov \\ Institute of Nuclear Physics,
Novosibirsk, Russia\thanks{permanent address, email:telnov@inp.nsk.su}
\\ and KEK, Japan \vspace{-3mm}} \date{}
\begin{document}
\newcommand{\M}{\mbox{m}}
\newcommand{\n}{\mbox{$n_f$}}
\newcommand{\EP}{\mbox{e$^+$}}
\newcommand{\EM}{\mbox{e$^-$}}
\newcommand{\EPEM}{\mbox{e$^+$e$^-$}}
\newcommand{\EMEM}{\mbox{e$^-$e$^-$}}
\newcommand{\GG}{\mbox{$\gamma\gamma$}}
\newcommand{\GE}{\mbox{$\gamma$e}}
\newcommand{\GP}{\mbox{$\gamma$e$^+$}}
\newcommand{\TEV}{\mbox{TeV}}
\newcommand{\GEV}{\mbox{GeV}}
\newcommand{\LGG}{\mbox{$L_{\gamma\gamma}$}}
\newcommand{\LGE}{\mbox{$L_{\gamma e}$}}
\newcommand{\LEE}{\mbox{$L_{ee}$}}
\newcommand{\WGG}{\mbox{$W_{\gamma\gamma}$}}
\newcommand{\EV}{\mbox{eV}}
\newcommand{\CM}{\mbox{cm}}
\newcommand{\MM}{\mbox{mm}}
\newcommand{\NM}{\mbox{nm}}
\newcommand{\MKM}{\mbox{$\mu$m}}
\newcommand{\SEC}{\mbox{s}}
\newcommand{\CMS}{\mbox{cm$^{-2}$s$^{-1}$}}
\newcommand{\MRAD}{\mbox{mrad}}
\newcommand{\IND}{\hspace*{\parindent}}
\newcommand{\E}{\mbox{$\epsilon$}}
\newcommand{\EN}{\mbox{$\epsilon_n$}}
\newcommand{\EI}{\mbox{$\epsilon_i$}}
\newcommand{\ENI}{\mbox{$\epsilon_{ni}$}}
\newcommand{\ENX}{\mbox{$\epsilon_{nx}$}}
\newcommand{\ENY}{\mbox{$\epsilon_{ny}$}}
\newcommand{\EX}{\mbox{$\epsilon_x$}}
\newcommand{\EY}{\mbox{$\epsilon_y$}}
\newcommand{\BI}{\mbox{$\beta_i$}}
\newcommand{\BX}{\mbox{$\beta_x$}}
\newcommand{\BY}{\mbox{$\beta_y$}}
\newcommand{\SX}{\mbox{$\sigma_x$}}
\newcommand{\SY}{\mbox{$\sigma_y$}}
\newcommand{\SZ}{\mbox{$\sigma_z$}}
\newcommand{\SI}{\mbox{$\sigma_i$}}
\newcommand{\SIP}{\mbox{$\sigma_i^{\prime}$}}
\maketitle

\vspace{-0.6cm}
{\large\it Abstract}
\vspace{2mm}

 It is very likely that in 3-4 years the construction of one or two
linear colliders with c.m.s energy up to 0.5--1.5 TeV will be
started. Besides \EPEM collisions, linear colliders give a unique
possibility to study \GG\ and \GE\ interactions at energies and
luminosities comparable to those in \EPEM\ collisions. High energy
photons for \GG\ and \GE\ collisions can be obtained using laser
backscattering.  These types of collisions considerably increase the
physics potential of linear colliders for relatively a small incremental
cost. This report briefly reviews the physics goals of \GG, \GE\
colliders and possible parameters of photon-photon colliders.

\vspace{5mm}
\hspace*{-3.5mm}{\large\bf 1 \ INTRODUCTION}
\vspace{0.3cm}

  The possibility of obtaining \GG,\GE\ colliding beams with high
energy and luminosity using Compton scattering of laser light on high
energy electrons at linear colliders (LC) has been considered since
1981 [1--5].
Possible parameters and physics potential of such colliders has been
discussed at many Workshops on LC  and at the 
Workshop on Gamma-Gamma colliders held in Berkeley in
1994~\cite{BERK}. Physics phenomena which can be studied in \GG,\GE\
collisions with high energies has been considered in many hundreds of
papers. This option is included now in the Conceptual Design Reports
of the NLC~\cite{NLC}, TESLA--SBLC~\cite{TESLA}, and JLC~\cite{JLC} linear
colliders. All these projects foresee a second interaction region for
\GG, \GE\ collisions.

  However, in our time of tight HEP budgets the physics community
needs a very clear understanding whether \GG,\GE\ collisions can
really give new physics information in addition to \EPEM\ collisions
that could justify an additional collider cost ($\sim$20\%, including
detector). In general, the physics at \EPEM\ and \GG,\GE\ colliders is
quite similar but complimentary, because cross sections depend
differrently on new unknown physics parameters. Roughly, the answer to
the previous question depends on the number of produced interesting
events (cross section $\times$ luminosity). If the statistics are
comparable, then \GG,\GE\ colliders should be built together with
\EPEM\ colliders without a doubt. In my opinion, this condition is
satisfied. Moreover, the beam collision effects allow more than one
order further increase of \GG\ luminosity though this will need
upgrading of the injector (decrease of the product of transverse beam
emittances).

\begin{figure}[!hbt]
\centering
\hspace*{-0.2cm} \epsfig{file=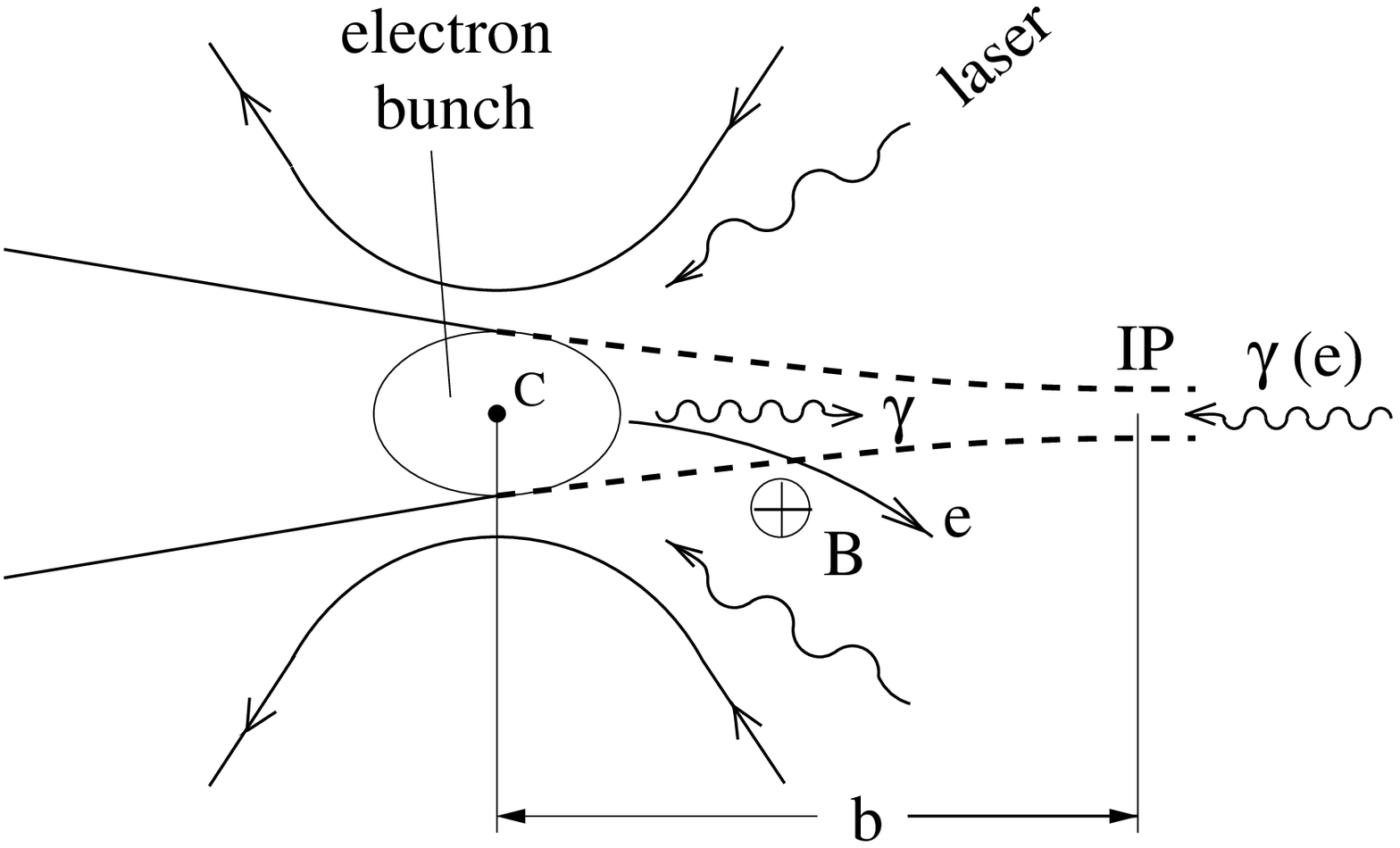,height=4.6cm,angle=-90} 
\vspace*{-0.15cm} 
\caption{Scheme of  \GG, \GE\ collider.}
\label{ris1}
\end{figure}

\begin{figure}[!hbt]
\centering
\vspace{-3mm}
\hspace*{-0.6cm} \epsfig{file=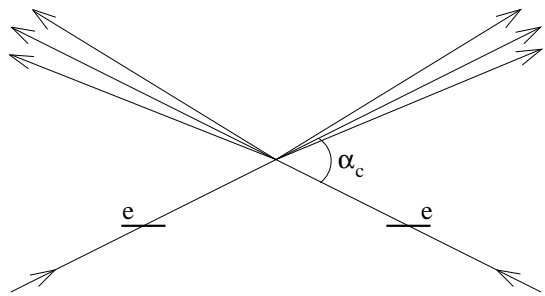,width=4cm,angle=-90}
\vspace*{-0.2cm} 
\caption{Crab-crossing scheme}
\vspace{-1mm}
\label{crab}
\end{figure}

The basic scheme of a photon collider is shown in Figs.~\ref{ris1} and
\ref{crab}.
Two electron beams after the final focus system are traveling toward
the interaction point (IP) and at a distance of about 0.1--1 cm from
the IP collide with the focused laser beams. Photons after  Compton
scattering have energies comparable with the energies of the initial
electrons and follow their direction (to the IP) with some small
additional angular spread of the order $1/\gamma $.  With reasonable
laser parameters one can ``convert'' most of the electrons into high
energy photons.  The luminosity of \GG, \GE\ collisions will be of the
same order of magnitude as the ``geometric'' luminosity of the basic
$ee$ beams. Luminosity distributions in \GG\ collisions have 
characteristic peaks near the maximum invariant masses with a typical
width about 10 \% (and a few times smaller in \GE\ collisions).  High
energy photons  can have various polarizations, which is very
advantageous for experiments.

    In the conversion region a photon with an  energy $\omega_0$  is
scattered on an  electron  with  an  energy $ E_0$   at  a  small
collision angle $\alpha_0$ (almost head-on). The energy of the scattered
photon $\omega$ depends on its angle $\vartheta$ with respect to the
motion of the incident electron as follows:
\vspace{-1mm}
\begin{equation}
\omega = \frac{\omega_m}{1+(\vartheta/\vartheta_0)^2};
\mbox{\hspace{0.1cm}} \omega_m=\frac{x}{x+1}E_0; \mbox{\hspace{0.1cm}}
\vartheta_0= \frac{mc^2}{E_0} \sqrt{x+1},
\end{equation}
$$x=\frac{4E_0 \omega_0 \cos^2 \alpha_0/2}{m^2c^4}
 \simeq 15.3\left[\frac{E_0}{TeV}\right]
\left[\frac{\omega_0}{eV}\right],$$
where $\omega_m$  is the maximum photon energy,

  For example: $E_0=300$ GeV, $\omega_0 =1.17$ eV (neodymium
glass laser) $\Rightarrow$ $x=5.37$ and $\omega/E_0 = 0.84$.  The value
$x=4.8$ is the threshold for \EPEM\ production in collision of the
high energy photon with a laser photon. Above this threshold
($x$ = 6--15) the yield of high energy photons will be lower by a factor
2--2.5.  Corresponding formulae and graphs can be found
elsewhere [2--5].

\vspace{5mm}
\hspace*{-3.5mm}{\large\bf 2 \ PHYSICS}
\vspace{0.3cm}

     The physics at high energy \GG, \GE\ colliders is very rich and
no less interesting than that in \EPEM\ or pp collisions:

1. The Higgs boson (which is thought to be responsible for the origin
of particle masses) will be produced at photon colliders as a single
resonance.  The cross section is proportional to the two-photon decay
width of the Higgs boson which is sensitive to all heavy charged
particles (even super-heavy) which get their mass via the Higgs
mechanism. In addition, some Higgs decay modes and its mass can be
measured at \GG\ colliders more precisely than in \EPEM\ collisions
due to larger production cross sections and the very sharp edge of the
luminosity spectrum.

2. Cross sections for production of charged scalar, lepton and top
  pairs in \GG\ collisions are larger than those in \EPEM\ collisions
  by a factor of approximately 5--10; for WW production this factor is
  even larger, about 10--20.

3. In \GE\ collisions, charged supersymmetric particles with masses
  higher than in \EPEM\ collisions can be produced (a heavy charged
  particle plus a light neutral); \GG\ collisions also provide higher
  accessible masses for particles which are produced as a single
  resonance in \GG\ collisions (such as the Higgs boson).

\vspace{2mm}

  The most intersting (expected) physics at next linear colliders is
the search for and study of the Higgs boson(s) and supersymmetric
particles. Photon colliders can make a considerable contribution to this
physics.

\begin{figure}[!htb]
\centering
\vspace*{-1.cm} 
\hspace*{-0.8cm} \epsfig{file=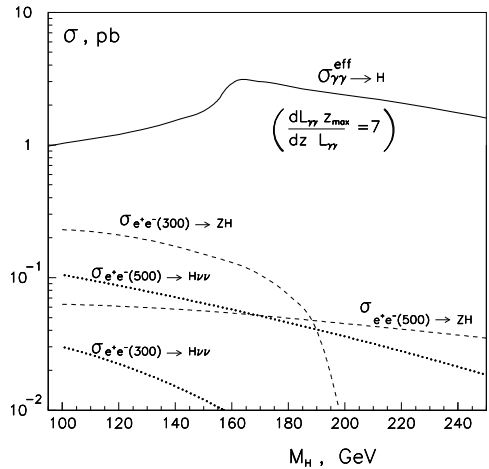,width=9.cm,angle=0} 
\vspace*{-1.4cm} 
\caption{Cross sections for the Standard model Higgs in \GG\ and
 \EPEM\ collisions.}
\label{cross}
\end{figure}

The mass of the Higgs most probably lies in the region of
100$<M_H<$300 GeV. The effective cross section is presented in
fig.~\ref{cross}~\cite{ee97}.  
Note that here \LGG\ is defined as the \GG\ luminosity at the high
energy luminosity peak ($z=\WGG/2E_e>0.65$ for $x=4.8$) with FWHM
about 15\%. The luminosity in this peak is approximately equal to
$0.25k^2\LEE(geom)$ ($k$ is the conversion coefficient).  For
comparison, in the same figure the cross sections of the Higgs
production in \EPEM\ collisions are shown. 

\begin{figure}[!htb]
\centering
\vspace*{-1.cm} 
\hspace*{-0.5cm} \epsfig{file=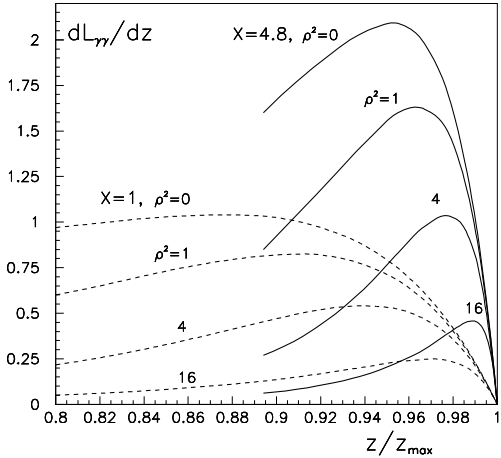,height=8.cm,angle=0}
\vspace*{-1.5cm}
\caption{ Shape of the \GG\ luminosity spectrum near the high energy edge.}
\vspace{-3.mm}
\label{edge}
\end{figure}

We see that for $M_H=$
120--250 GeV the effective cross section in \GG\ collisions is larger
than that in \EPEM\ collisions by a factor of about 6--30!  If the
Higgs is light enough, its width is much less than the energy spread in
\GG\ collisions. It can be detected as a peak in the invariant mass
distribution or can be seached by energy scanning using the very sharp
edge of luminosity distribution (see fig.\ref{edge}).
  Observation of a sharp step in the visible cross section will imply
narrow resonance production with subsequent decay in the considered
channel. This method is very attractive for study of the Higgs in the
$\tau\tau$ decay mode where direct reconstruction is impossible due to
undetected neutrinos while it can be seen as a step in visible cross
section for events consisting of two low multiplicity collinear jets.
The total number of events in the main decay channels $H \to b\bar b,
WW(W^*), ZZ(Z^*)$ will be several thousands for a typical integrated
luminosity of 10 fb$^-1$ \cite{ee97}. The scanning method also allows the measurement of the Higgs mass with high precision.
   
\begin{figure}[!thb]
\centering
\vspace*{-0.9cm}
\hspace*{-0.5cm} \epsfig{file=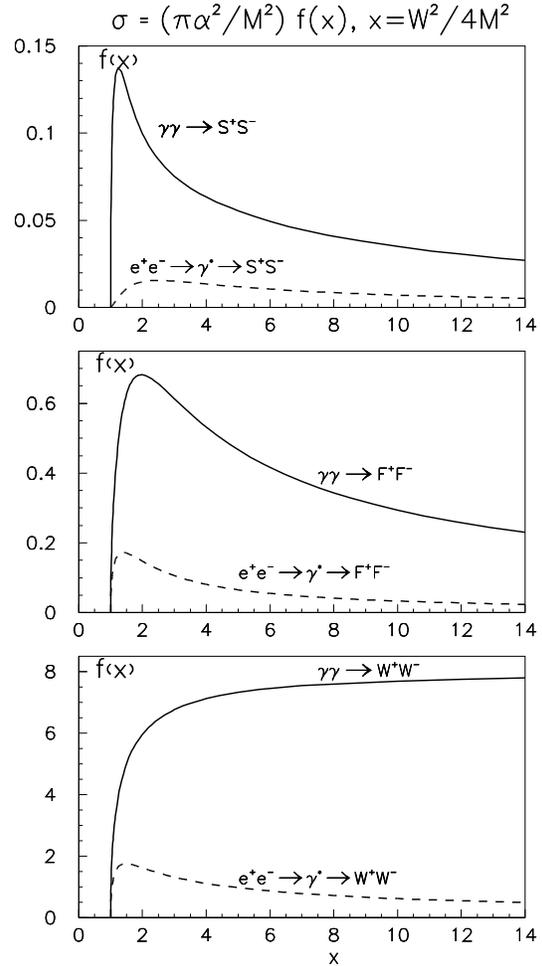,height=14.5cm,width=8.9cm}
\vspace*{-1.5cm}
\caption{ Cross sections for charged pair production in
\EPEM\ and \GG\ collisions. S (scalars), F (fermions), W (W-bosons); M
is particle mass, W is invariant mass (c.m.s. energy of colliding beams)}
\vspace{-3.mm}
\label{fig16a}
\end{figure}

\begin{figure}[!thb]
\centering
\vspace*{-0.7cm}
\hspace*{-0.5cm} \epsfig{file=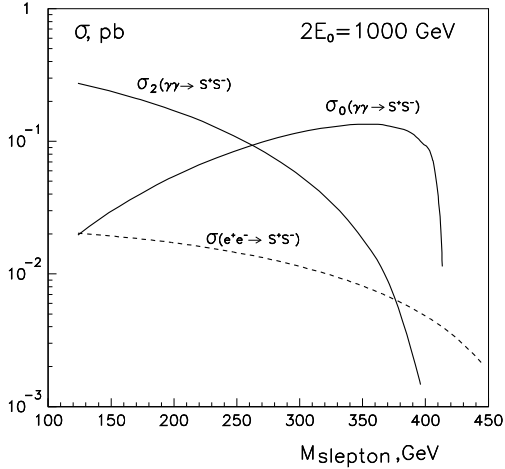,width=9.2cm}
\vspace*{-1.6cm}
\caption{ Cross sections for charged boson production in \EPEM\ and
\GG\ collisions at $2E_0$ = 1 TeV collider (in \GG\ collision
$W_{max}\approx 0.82$ GeV ($x=4.6$)); $\sigma_0$ and $\sigma_2$ correspond to
the total \GG\ helicity 0 and 2.}
\vspace{-2mm}
\label{crossel}
\end{figure}

   The second example is a charged pair production. The corresponding
cross sections in unpolarized \GG\ and \EPEM\ collisions are shown in
Fig.\ref{fig16a}.  
One can see that in \GG\ collisions the cross sections are much
larger, by at least  a factor 5 for scalars and fermions and by about 
one order in WW channel. The cross section of scalar pair production
(sleptons, for example) in collision of polarized photons is shown
in fig.\ref{crossel}.
One can see that for heavy scalars
the cross section in collisions of polarized photons is higher than that
in \EPEM\ collisions by a factor of 10--20.  

\vspace{5mm}
\hspace*{-3.5mm}{\large\bf 3 \ LUMINOSITY OF \GG\ COLLIDERS \\
\hspace*{9mm} IN CURRENT DESIGNS}
\vspace{0.3cm}

\hspace*{-3.5mm}{\large\it 3.1  \ 0.5--1 TeV colliders}
\vspace{0.2cm}

    Below some results of simulation of \GG\ collisions at
TESLA, ILC (converged NLC and JLC) and CLIC are presented. Beam parameters were
taken the same as in \EPEM\ collisions with the exception of horizontal
beta function at IP, which is taken conservatively equal to 2 mm for
all cases. In \GG\ collisions the beamstrahlung is absent and the
horizontal size can be made much smaller than that in \EPEM\
collisions. Minimum $\beta_x$ is determined by the Oide effect (radiation
in quads) which is included in the simulation code and also by
technical problems connected with the chromatic corrections in both
transverse directions -- the limit here is not so
clear. The conversion point(CP) is situated at the distance
$b=\gamma\sigma_y$. It is assumed that electron beams have 85\% longitudinal
polarization and laser photons have 100\% circular polarization.

 The simulation code~\cite{TEL95} takes into account all important
processes: linear Compton scattering with all polarization effects,
beamstrahlung (without polarization effects), coherent pair creation and
interaction between charged particles.

{\setlength{\tabcolsep}{0.1mm}
{\footnotesize
\begin{table}[!hbtp]
\caption{Parameters of  \GG\ colliders based on Tesla(T), ILC(I)
and CLIC(C).}
\vspace{-0.2cm}
\begin{center}
\hspace*{0.5mm}\begin{tabular}{c c c c c c c} \hline
 & T(500) & I(500) & C(500) \ &
  T(800) & I(1000) &  C(1000) 
                                                   \\ \hline \hline 
\multicolumn{7}{c}{ no deflection, $b=\gamma \sigma_y$, $x=4.6$} \\ \hline
$N/10^{10}$& 2. & 0.95 & 0.4 & 1.4 & 0.95 & 0.4 \\  
$\sigma_{z}$, mm& 0.4 & 0.12 & 0.05 & 0.3 & 0.12 & 0.05 \\  
$f_{rep}\times n_b$, kHz& 15 & 11.4 &30.1& 13.5 & 11.4 & 26.6 \\
$\gamma \epsilon_{x,y}/10^{-6}$,m$\cdot$rad & $10/0.03$ & $5/0.1$ & 
$1.9/0.1$&  $8/0.01$ & $5/0.1$ & $1.5/0.1$ \\
$\beta_{x,y}$,mm at IP& $2/0.4$ & $2/0.12$ & $2/0.1$ &
$2/0.3$& $2/0.16$ & $2/0.1$ \\
$\sigma_{x,y}$,nm& $200/5$ & $140/5$ & $88/4.5$ & 
$140/2$ & $100/4$ & $55/3.2$ \\  
b, mm & 2.4 & 2.4 & 2.2 & 1.5 & 4 & 3.1 \\
$L(geom),\,\,\,  10^{33}$& 48 & 12 & 10 & 75 & 20 & 19.5\\  
$\LGG (z>0.65), 10^{33} $ & 4.5 & 1.1 & 1.05 & 7.2 & 1.75 & 1.8 \\
$\LGE (z>0.65), 10^{33}$ & 6.6 & 2.6 & 2.8 & 8  & 4.2 & 4.6 \\
$\LEE, 10^{33}$ & 1.2 & 1.2 & 1.6 & 1.1 & 1.8 & 2.3 \\
$\theta_x/\theta{_y},_{max}$, mrad ~ & 5.8/6.5 & 6.5/6.9 & 6/7& 
 4.6/5 & 4.6/5.3 & 4.6/5.5 \\ \hline
\vspace{-7.mm}
\end{tabular}
\end{center}
\label{table1}
\end{table}
}} 

We see that \GG\ luminosity in the hard part of the spectrum is $\LGG
  (z>0.65)\sim 0.1L(geom)\sim (1/6)L_{\EPEM}$. 

Beside \GG\ collisions, there is considerable \GE\ luminosity which
adds some background (\EPEM pairs in  vertex detectors), but on the
other hand, it is possible to study \GE\ interactions simultaneosly
with \GG\ collisions. Optimization of \GG\ and \GE\ luminosities was
considered in  refs.\cite{TEL90},\cite{TEL95},\cite{TESLA}.
  
\begin{figure}[!htb]
\centering
\vspace*{-2.cm} 
\hspace*{-0.8cm} \epsfig{file=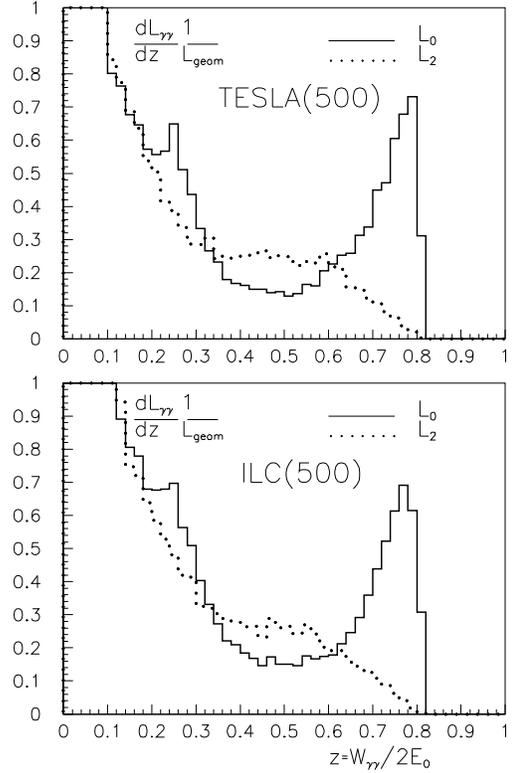,width=9.55cm,angle=0} 
\vspace*{-1.9cm} 
\caption{Luminosity spectra for \GG\ collisions for collider
parameters presented in table 1. Solid line for total helicity of two
photons 0 and dotted line for total helicity 2.}
\vspace{-3.5mm}
\label{TeslaIlc}
\end{figure}

   The normalized \GG\ luminosity spectra for a 0.5 TeV TESLA and ILC colliders
colliders are  shown in Fig.\ref{TeslaIlc}. 
The luminosity spectrum is decomposed into two parts, with total
helicity of two photons 0 and 2. We see that in the high energy part
of the luminosity spectra photons have high degree of polarization,
which is very important for many experiments.  In addition to the high
energy peak, there is a factor 5--8 larger low energy luminosity. It is
produced by photons after multiple Compton scatterings and
beamstrahlung photons. Fortunately, these event have large boost and
can be easily distinguished from the central high energy events.

\begin{figure}[!htb]
\centering
\vspace*{-2.cm} 
\hspace*{-0.8cm} \epsfig{file=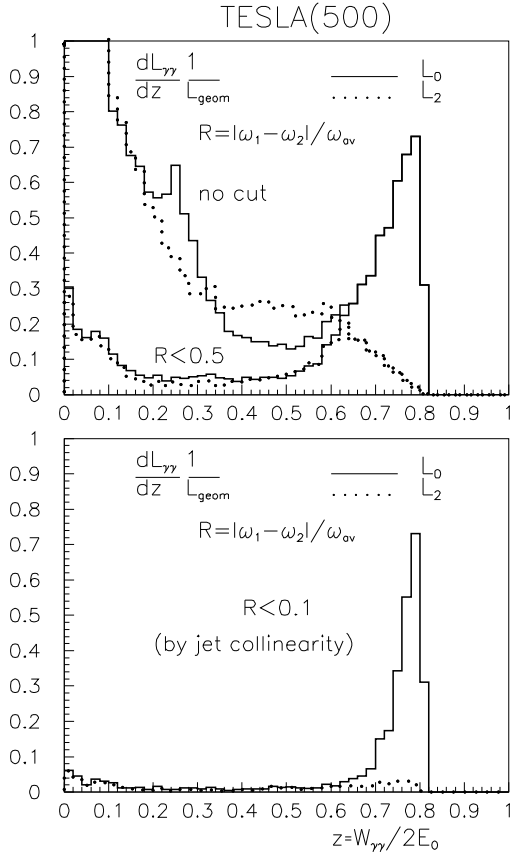,width=9.8cm,angle=0} 
\vspace*{-2.3cm} 
\caption{Luminosity spectra for \GG\ collisions at TESLA(500) with various cut
on the relative difference of the photon energy. See comments in the text.}
\vspace*{-3mm} 
\label{TeslaR}
\end{figure}

Fig.\ref{TeslaR} (upper) shows the same spectrum 
with an additional cut on the longitudinal momentum of the produced
system  which suppresses low energy luminosity to a
neglegible level. 
 Fig.\ref{TeslaR} (lower) shows the same spectrum with a stronger cut
on the longitudinal momentum. In this case the spectrum has a nice peak with
FWHM about 7.5\%. Of course, such procedure is somewhat artificial
because instead of such cuts one can directly selects events with high
invariant masses, the minimum width of the invariant mass distribution
depends only on the detector resolution. However, there are very
important examples when one can obtain a ``collider resolution''
somewhat better than the detector resolution, such as the case of only
two jets in the event when one can restrict the longitudinal momentum
of the produced system using the acollinearity angle between jets ($H\to
b\bar b, \tau\tau$, for example).

Detailed background studies~\cite{TESLA} show that the low energy part
($z<0.6$) of the \GG\ luminosity increases hadronic background in the
detector by only a factor of 2. In the scheme with deflection of
electrons by the magnetic field of a small magnet between IP and CP it is
possible to suppress the low energy part of the \GG\ luminosity by several
times ---  however, this approach is more complicated technically and the
maximum \GG\ luminosity is smaller than that without deflection (due to the
larger space between CP and IP).

\vspace{0.3cm}
\hspace*{-3.5mm}{\large\it 3.2  \ \GG\ collider for low mass Higgs}
\vspace{0.2cm}

  It is very possible that the Higgs boson has a mass in the region
115-150 GeV as predicted in some theories. It is of interest to consider
possible parameters of a \GG\ collider based on TESLA and ILC at these
energies. Two variants were considered for H(130): 1) the ``Compton''
parameter $x$ is fixed near the threshold of \EPEM\ creation ($x
\approx 4.6$), which corresponds to $\lambda \sim 325$ nm and $E_0=79$
GeV; 2) the laser is the same as for $2E_0 = 500$ GeV colliders, namely
a Nd:glass laser with $\lambda=1.06\;\mu m$, which corresponds to $x=1.8$
and $E_0=100$ GeV. All other beam parameters are taken the same as for
$2E_0 = 500$ GeV (see Table~\ref{table1}). Results of simulation for these
two cases are shown in Table \ref{table2} (TESLA and ILC) and in
Fig.\ref{Higgs130} (TESLA).
{\setlength{\tabcolsep}{1mm}
{\footnotesize
\begin{table}[!hbt]
\caption{Parameters of the \GG\ colliders for 
Higgs(130) at TESLA(T)and ILC(I).}
\vspace{-0.2cm}
\begin{center}
\hspace*{-4mm}\begin{tabular}{c c c  c c} \hline
 & T(2x100) & I(2x100) & T(2x79) & I(2x79) 
                                                   \\ \hline \hline 
\multicolumn{5}{c}{\mbox{\hspace*{2.6cm}} $x=1.8$ \mbox{\hspace*{1.4cm}}\
 $x=4.6$ } \\ \hline
$\sigma_{x,y}$,nm& $320/7.8$ & $230/7.8$ & $360/8.8$ & $250/8.8$ \\  
b, mm & 1.5 & 1.5 & 1.4 & 1.4 \\
$L(geom),\,\,\,  10^{33}$& 19 & 4.6 &  15 & 3.7 \\  
$\LGG (z/z_m>0.8),10^{33}$ & 1.55 & 0.37 & 1.45 & 0.35 \\
$\LGE (z/z_m>0.8),10^{33}$ & 3. & 1.45 & 1.7 & 0.83 \\
$\theta_x/\theta{_y},_{max}$, mrad ~ & 5.2/6.2 & 5.2/7 & $\sim$ 10/12 & 
 $\sim$ 10/12 \\ \hline
\end{tabular}
\end{center}
\label{table2}
\vspace{-3mm}
\end{table}
} 
\begin{figure}[!htb]
\centering
\vspace*{-1.3cm} 
\hspace*{-0.8cm} \epsfig{file=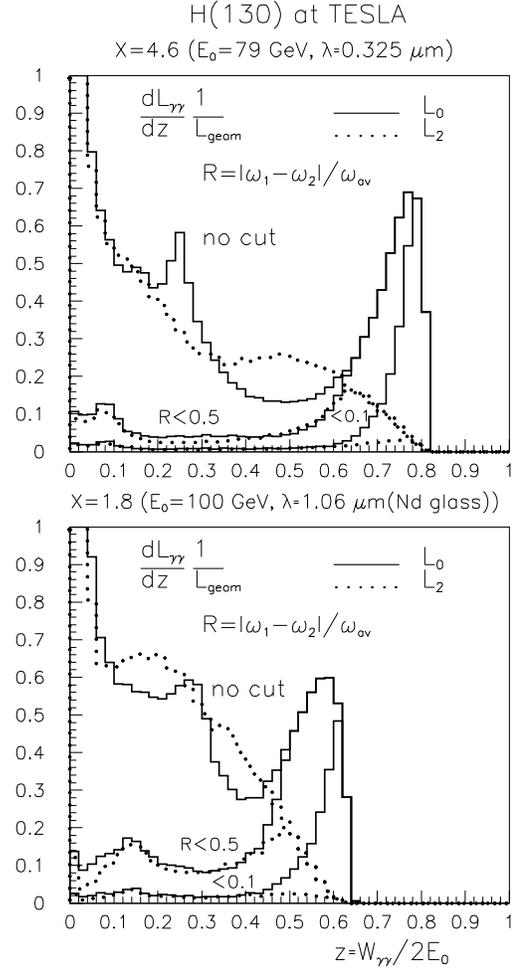,width=10.cm,angle=0} 
\vspace*{-2.3cm} 
\caption{Luminosity spectra of \GG\ collision of `low' energy \GG\ collider
(TESLA beam parameters) for study of the Higgs with a mass
$M_H=130$ GeV, upper figure for $x=4.8$ and lower for $x=1.8$
(the same laser as for $2E_0=500$ GeV).}
\vspace{-1mm}
\label{Higgs130}
\end{figure} 
Comparing these two variants we see that peak luminosities ($dL/dz$)
 are approximately the same (note that $L_{geom}$ at $x=1.8$ is larger
 by a factor 1.26 due to larger energy), and the ratio $L_0/L_2$ is
 also the same (the Higgs is produced by $L_0$ and main backgrounds
 comes from $L_2$). The only difference is that the slope of the
 luminosity at $z$ near $z_{max}$ is larger for $x=4.6$ by a factor
 2.3 (important for measurement of Higgs mass).  Also the maximum
 disruption angle at $x=4.8$ is larger by a factor 1.7. This angle
 determines the minimum crab crossing angle at the interaction point
 (see Fig.\ref{crab}). For 2E=500 GeV colliders the maximum disruption
 angle is 10 mrad (safely) and the crab-crossing angle $\alpha_c=30$
 mrad. If we keep $x=const$, then at $2E_0=2\times79$ GeV the maximum
 disruption angle will be larger by a factor $\sqrt{250/79}=1.8$,
 which already introduces some problems.
  
  From this consideration we can conclude that one can use the same
Nd:glass laser at all energies below $2E_0 \sim$ 500 GeV.

\vspace{5mm}
\hspace*{-3.5mm}{\large\bf 4 \ ULTIMATE \GG\ LUMINOSITY }
\vspace{0.3cm}

   The \GG\ luminosities in the current projects are determined by
the ``geometric'' luminosity of the electron beams. The only collision effect
restricting the maximum value of the \GG\ luminosity is  coherent
pair creation when the high energy photon is converted into an \EPEM\ pair in
the field of the opposing electron beam~\cite{CHEN},\cite{TEL90},\cite{TEL95}.
Having electron beams with smaller emittances one can obtain much higher
\GG\ luminosity~\cite{TSB2}. Fig.\ref{sigmax} 
\begin{figure}[!htb]
\centering
\vspace*{-0.9cm} 
\hspace*{-0.8cm} \epsfig{file=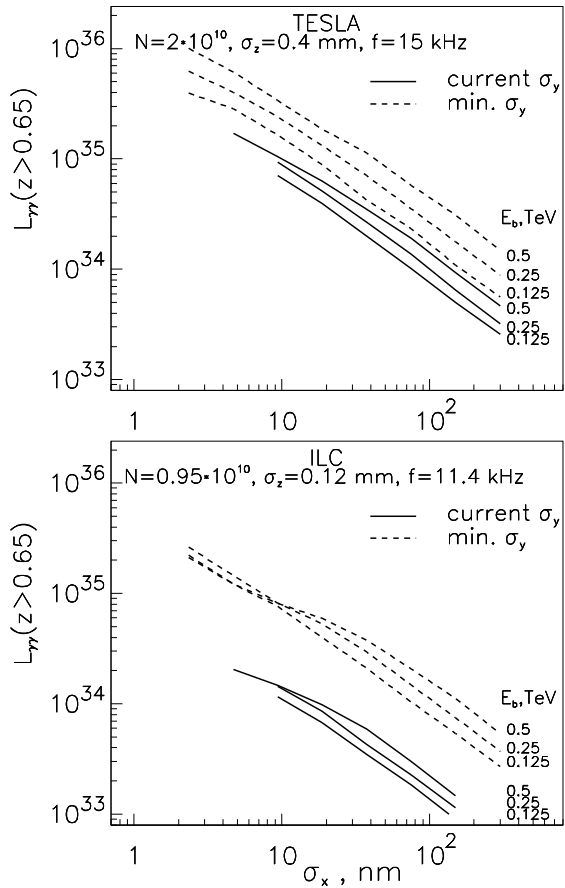,width=9.cm,angle=0} 
\vspace*{-1.6cm} 
\caption{Dependence of \GG\ luminosity on the horizontal beam size for
TESLA and ILC at various energies. Solid curves for nominal vertical
beam size, dashed curves for minimum vertical beam size determined by the 
minimum distance between the conversion and interaction points.}
\vspace{0mm}
\label{sigmax}
\end{figure} 
shows dependence of the
\GG\ luminosity on the horizontal beam size. Solid curves
correspond to the case where the vertical emittance is the same as in
TESLA(500), ILC(500) projects (see Table \ref{table1}). Dashed curves
represent the case where the vertical beam sizes are as small as
possible and are determined only by the minimum distance between the
interaction and conversion points ($\sigma_y \sim b/\gamma$), where
$b_{min}\sim 3\sigma_z + 0.08E[\TEV]$ cm. The second term is the
half length of the conversion region determined by nonlinear effects
in Compton scattering.

   One can see that all curves follow their natural behavior:
$\L\propto 1/\sigma_x$, with the exception of ILC at $2E_0=1$ GeV where
the effect of coherent pair creation is seen. This means that at the
same colliders the \GG\ luminosity can be increased almost by two
orders. Even with one order improvement, the number of ``interesting''
events (the Higgs, charged pairs) at photon colliders will be larger
than that in \EPEM\ collisions by more that one order. This is a nice
goal and motivation for photon colliders.

   There are several ways of decreasing transverse beam emittances
(their product): optimization of storage rings (with long wigglers)
and low-emittance guns (with merging many beams with low emittances).  Here
 progress is certainly possible. Morover, there is one method which
allows further decrease of beam cross sections by two orders in
comparison with current designs. It is laser
cooling~[13-14].  In this method the electron beam
at an energy of about 5 GeV is collided 1--2 times with a powerful laser
flash, losing in each collision a large fraction ($\sim 90\%$) of its
energy to radiation, with reacceleration between cooling
sections. The physics of the cooling proccess is the same as in a wiggler.
The required flash energy is about 10 Joules. This scheme can be
realized, in principle, already.  If this upgrading of luminosity is
to be done in 15 years from now then, certainly, it will be no
problem. For example, the peak power of lasers  increased during the last
ten years by 4 orders of magnitude. All laser technologies required
for photon colliders are being developed actively now for other
applications.

\vspace{5mm}
\hspace*{-3.5mm}{\large\bf 5 \ CONCLUSION}
\vspace{2mm}
    
   Prospects of photon colliders for particle physics are great; the physics
community should not miss this unique possibility.     

\newpage
\hspace*{-3.5mm}{\large\bf Acknowledgements}
\vspace{0.3cm} 

I would like to thank all my colleagues for joint work, useful
discussions and support of this direction. Special thanks to A.Sessler
for his strong promotion of gamma-gamma colliders~\cite{BERK},\cite{PToday}.
\vspace{-3mm}

\end{document}